\documentclass[aip,jmp,amsmath,amssymb,reprint]{revtex4-1}
\usepackage{graphicx}
\usepackage{dcolumn}
\usepackage{bm}
\begin{document}

\preprint{AIP/123-QED}

\title[Sample title]{Efficiency calculation of thermoelectric generator using temperature dependent material's properties}
\author{Kumar Gaurav}
\email{kumargauravmenit@gmail.com}
 \altaffiliation[also at]{ School of Engineering. Indian Institute Of Technology Mandi, Kamand, Himachal Pradesh, India, 175005}
\author{Sudhir K Pandey}
 \email{sudhir@iitmandi.ac.in}
\affiliation{ School of Engineering, Indian Institute of Technology Mandi, Kamand, Himachal Pradesh, India, 175005} 
\date{\today}
\begin{abstract}
Accurate measurement of efficiency for thermoelectric generator (TEG) is of great importance for materials research and development. Approximately all the parameters of a material are temperature dependent, so we can't directly apply the $\eta_\text{max}$ formula for efficiency calculation in the large temperature range. To overcome that problem, we tried to calculate the efficiency of TEG by dividing large working temperature range into a number of small temperature difference. The aim is to make temperature dependent parameter to be constant for that small temperature range. Using maximum individual efficiency of each segment obtained by $\eta_\text{max}$ in the equation of $\eta_\text{overall}$, which gives overall efficiency. The $\eta_\text{overall}$ of TEG using $Bi_2Te_3$ and $TAGS$ as thermoelectric materials come out to be 7.1\% and 8.94\%, respectively, which is close to experimental results. For the high-temperature region, we have used $SiGe$ material in TEG and found out $\eta_\text{overall}=3.5\%$. The cumulative efficiency obtained by keeping one end temperature fixed with another end varying can be applied in real life application, i.e. automobile sector. The present work provides a simple way for the design engineers to calculate the efficiency of TEG by using the temperature dependent materials parameters like thermal conductivity, electrical conductivity, and Seebeck coefficient on which $z\bar{T}$ depends.\\
keywords: waste heat recovery; thermoelectric generator (TEG); power generation; automobile; efficiency;
\end{abstract}
\bigskip]
\maketitle
\thispagestyle{plain}
\pagestyle{plain}
\section{INTRODUCTION}
Nowadays, due to heavy industrialization and increased demand of power for day to day life applications, we need to supply an enormous amount of energy. Mostly energy is being produced by coal, petroleum, hydro, solar, wind, nuclear, etc., where globally around 65\% of electrical energy is being produced by coal and petroleum only. We can see the extent of our dependency on non-renewable natural resources, where stocks are limited. If we keep on using these non-renewable resources at the same pace, currently available stocks are going to be exhausted in next 50 years. If we take an example of India, where $\sim$70\% of power is being produced by thermal plant\cite{india}, which fully depends on the limited resources of fossil fuels. These thermal plants operate at around 32\% efficiency, which means approximately 68\% of energy is being wasted out. The same situation is in the automotive sector, where around two third of energy is being wasted out\cite{tervo}. So, if there is any possibility to extract some useful energy from that waste heat, then it will help us to reduce our dependencies over coal and petroleum to some extent\cite{leblanc}. For extracting energy from waste heat, we have thermoelectric generators made of thermoelectric materials are the best candidate. TEG has many novel qualities such as it is eco-friendly, having no moving parts that's why it needs less maintenance.\par In spite of the above-mentioned novelty of TEG, there are certain challenges in developing a good TEG for a particular application and one of them is efficiency. As we know that the TEG works on the principle\cite{rodger} of temperature difference between two ends of the sample, which creates a potential difference between hot and cold ends of the sample. Consequently, electrical energy is being produced. In general, efficiency is defined as:-
\begin{equation}
\eta= \frac{\text{Net power output}}{\text{Heat Input}}
\end{equation}
The maximum efficiency\cite{sherman} of a TEG using thermoelectric material is:
\begin{equation}
\centering
\eta_\text{{max}}={\frac{T_{h}-T_{c}}{T_{h}}}\frac{\sqrt{1+z\bar{T}}-1}{\sqrt{1+z\bar{T}}+\frac{T_{c}}{T_{h}}}
\end{equation}
where $\bar{T}=\frac{T_{h}+T_{c}}{2}$, $T_{h}$, and $T_{c}$ are average temperature, source temperature, and sink temperature, respectively. $z\bar{T}$ is called as figure of merit\cite{rowe}, where $z\bar{T}$ is defined as:
\begin{equation}
z\bar{T}=\frac{S^{2}\sigma}{\lambda}\bar{T}
\end{equation} 
In Eq.(3) S, $\sigma$, $\lambda$ are Seebeck coefficient, electrical conductivity, and thermal conductivity, respectively of thermoelectric material used for making TEG\cite{taylor}. Above formula for $\eta_\text{{max}}$ gives us criterion for selecting a material for fabricating TEG, which suggests that higher the $z\bar{T}$, higher will be the $\eta_\text{{max}}$ for a fixed value of $T_{h}$ and $T_{c}$. Here $z\bar{T}$ appears as constant, however $z\bar{T}$ has temperature dependence\cite{tritt} as can be seen from Eq.(3). Although it appears to be simple, but the use of Eq.(2) becomes tricky if $z\bar{T}$ varies appreciably in a given temperature range because we can't take a fixed value of $z\bar{T}$ for broad temperature range. Hence, we can't apply Eq.(2) directly.\par To overcome the problem regarding temperature dependence aspect of $z\bar{T}$, we have adopted a technique of segmentation of sample. Here segmentation of sample refers to dividing the entire working temperature range into a number of small-small sections. These small-small temperature segments will make $z\bar{T}$ as temperature independent identities for that small temperature range. After getting the individual efficiency of each section, we can apply thermodynamic formula for the efficiency calculation of entire sample.\par Firstly, we took ${Bi_2Te_3}$ under consideration for a temperature range of 310 K to 590 K. Using $z\bar{T}$ for each small segmented section for respective temperature range, we found out overall efficiency as 7.1\%. Simultaneously, we found out the efficiency of TEG using $TAGS$ as 8.95\% and $SiGe$ as 3.5\% for a temperature range of 580 K - 830 K and 870 K - 1260 K, respectively. For validating our results with experimentally obtained results, we also calculated the cumulative efficiency of TEG keeping fixed cold end temperature with varying hot end temperature and successfully reached to reasonable results. Applicability of our method and obtained results will be justified only when TEG is being installed in some real life application like automobile\cite{snyder2008}. Usually, in automobile higher temperature remains as fixed. For that, we have calculated temperature dependent efficiency of $Bi_2Te_3$, $TAGS$, and $SiGe$ keeping higher temperature fixed and varying lower temperature as per their favorable operating temperature range.

\section{METHODOLOGY} 
In order to calculate the $\eta_\text{{max}}$ from the temperature dependence data of $z\bar{T}$, we have divided working temperature range into "n" number of segments as shown in Fig.1, where $T_{h}$ and $T_{c}$ denote the temperature of hot and cold ends, respectively. For that small segmented temperature difference, we can assume fixed value of $z\bar{T}$. This temperature difference of each segment can be decided by looking at the temperature dependent behavior of $z\bar{T}$ for the sample. The number of slicing will depend on the temperature range of each sliced part. If $\Delta{T}$ is a temperature difference across each slice which gives an almost constant value of $z\bar{T}$ in that region then the number of segments will be $\frac{T_{h}-T_{c}}{\Delta{T}}$.
\begin{figure}[ht!]
\centering
\includegraphics[width=1.0in]{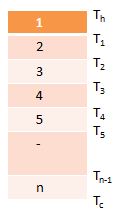}
\caption{schematic of  segmenting of the sample, considering 1-D flow of heat.}
\end{figure}
\\First, consider segment 1, for which $\Delta{T}= T_{h}-T_{1}$, where $T_{h}$ and $T_{1}$ are defined as $T_{h}= \bar{T}+\frac{\Delta T}{2}$, $T_{1}=\bar{T}-\frac{\Delta T}{2}$ and $z\bar{T}$ is the figure of merit for this segment. By putting $z\bar{T}$, $T_{h}$ and $T_{1}$ in Eq.(2), we get the efficiency of segment (1), i.e,
\begin{equation}
\centering
\eta_{1}= {\frac{\Delta T}{\bar{T}+\frac{\Delta T}{2}}}{\frac{\sqrt{1+z\bar{T}}-1}{\sqrt{1+z\bar{T}}+\frac{\bar{T}-\frac{\Delta T}{2}}{\bar{T}+\frac{\Delta T}{2}}}}
\end{equation}
Similarly, by using the above procedure for each segment, we can get the efficiency of all individual segment, which are $\eta_{1}$, $\eta_{2}$, $\eta_{3}$,....., and $\eta_{n}$ for $1^{st}$, $2^{nd}$, $3^{rd}$,...., and $n^{th}$ segments, respectively. Considering all the individual segmented efficiency that are in series, we can apply the fundamental law of thermodynamics\cite{okamoto} for efficiency calculation of segmented system. There is a fundamental difference between the cascaded and segmented system, in cascaded system output is taken for each sliced part but in segmented system net output power is being taken between both extreme ends of sample\cite{chen},\cite{zhang}. Which gives overall efficiency of TEG as\cite{ybarrondo}
\begin{equation}
\centering
\eta_\text{overall}=1-(1-\eta_{1})(1-\eta_{2})(1-\eta_{3})......(1-\eta_{n})
\end{equation}For using Eq.(5), we have taken a necessary assumption of not considering any type of losses because of conduction, convection and radiation from the lateral surface as well as interface contact.
\section{Result and Discussion}
It is clear from the above-discussed method that one can calculate the efficiency of a TEG working in a given temperature range where $z\bar{T}$ has temperature dependence. Now we want to verify the accuracy of the calculated efficiency using above method by comparing with experimentally observed efficiency. In this work, we are focusing only on $Bi_{2}Te_{3}$, $TAGS$, and $SiGe$ as thermoelectric materials to be used in TEG. Firstly, we want to apply the method for $Bi_{2}Te_{3}$ in the temperature range of 310 K to 585 K. Here, we have considered $\Delta{T} = 10 K$ and used the temperature dependent $z\bar{T}$ data from the literature\cite{snyder2004}. Segmentation of entire sample is done, which gives total 28 segments. For the first segment $\bar{T} = 580 K$. Using the value of $\bar{T}$, $\Delta{T}$ and $z\bar{T}=0.125$ at 580 K in Eq.4, we get $\eta_{1}=0.000509\%$. Applying similar procedure, we got the efficiency of the consecutive layers which are $\eta_{2}=0.000581\%$,......, and $\eta_{28}=0.004472\%$. Using all these segmented efficiencies in Eq.(5), we got $\eta_\text{overall}= 7.1\%$. The result obtained through this approach is close to the experimental value\cite{sano} of $7\%$ in the temperature range of 310 K - 585 K. The total number of segments considered here is 25, which gives the overall calculated efficiency $8.95\%$ in the temperature range of 830 K - 580 K. This is also in good agreement with the experimental result\cite{salzgeber}. Similarly, for higher temperature range $\sim$ 1250 K $SiGe$ is taken into consideration. We have calculated the efficiency TEG for $SiGe$ in the temperature range of 870 K - 1260 K. In this temperature range, a total number of segment comes out to be 39. The calculated efficiency comes out $3.5\%$. In the absence of experimental efficiency, we could not compare the calculated result with the experiment. Since, the calculated efficiency of TEG using $SiGe$ is $3.5\%$, which is significantly low, hence our result suggests that $SiGe$ is not a good material for TEG in this temperature range. Theoretically, we have calculated the efficiency of TEG and for comparison with experimental results we have tabulated our results in Table(1). 
\begin{table}
\centering
\caption{Validation of our theoretically calculated result with experimentally observed results.}
\begin{tabular}{|p{1.5cm}||p{0.8cm}|p{0.8cm}|p{1.8cm}|p{2.2cm}|}
\hline
$\text{Materials}$&\text{T$_{h}$ K}&\text{T$_{c}$ K}&\text{Calculated\%}&\text{Experimental\%} \\ \hline
$Bi_{2}Te_{3}$ &590 &305 &7.10 &7 \\ \hline
$TAGS$ &830 &580 &8.94 &8.9 \\ \hline
$SiGe$ &1260 &870 &3.50&\text{---} \\ \hline
\end{tabular}
\end{table}
\par It is clear from the above discussion that either by using the temperature dependent data of $z\bar{T}$ or the temperature dependent S, $\sigma$, $\lambda$ data, one can calculate the efficiency of a TEG operating in a particular temperature range. This method can also be used to calculate the temperature dependent efficiency of TEG when the temperature of one end is fixed and that of another end is varying. We have calculated the temperature dependent efficiency for $Bi_{2}Te_{3}$ when temperature of cold end is fixed at 310 K and the temperature of hot end is varying in the step of 10 K, up to 590 K. The efficiency when higher temperature is T K and fixed cold end temperature $T_{c}$, can be written as:
\begin{equation}
 \eta_{T}=1-(1-\eta_{n})(1-\eta_{n-1})(1-  \eta_{n-2})......(1-\eta_{i})
\end{equation} where $T=T_{c}+\Delta{T}(n-i+1)$ and $\Delta{T}$ is the temperature difference across each segment. Temperature dependent efficiency obtained through this formula is plotted in Fig.2. This temperature dependent efficiency can be validated with work done by Salzgeber et al. \cite{salzgeber}. They have also reported temperature dependent efficiency of TEG made up of $Bi_2Te_3$ for a fixed $T_{c} = 310 K$.
\begin{figure}[ht]
\centering
\includegraphics[width=3.5 in]{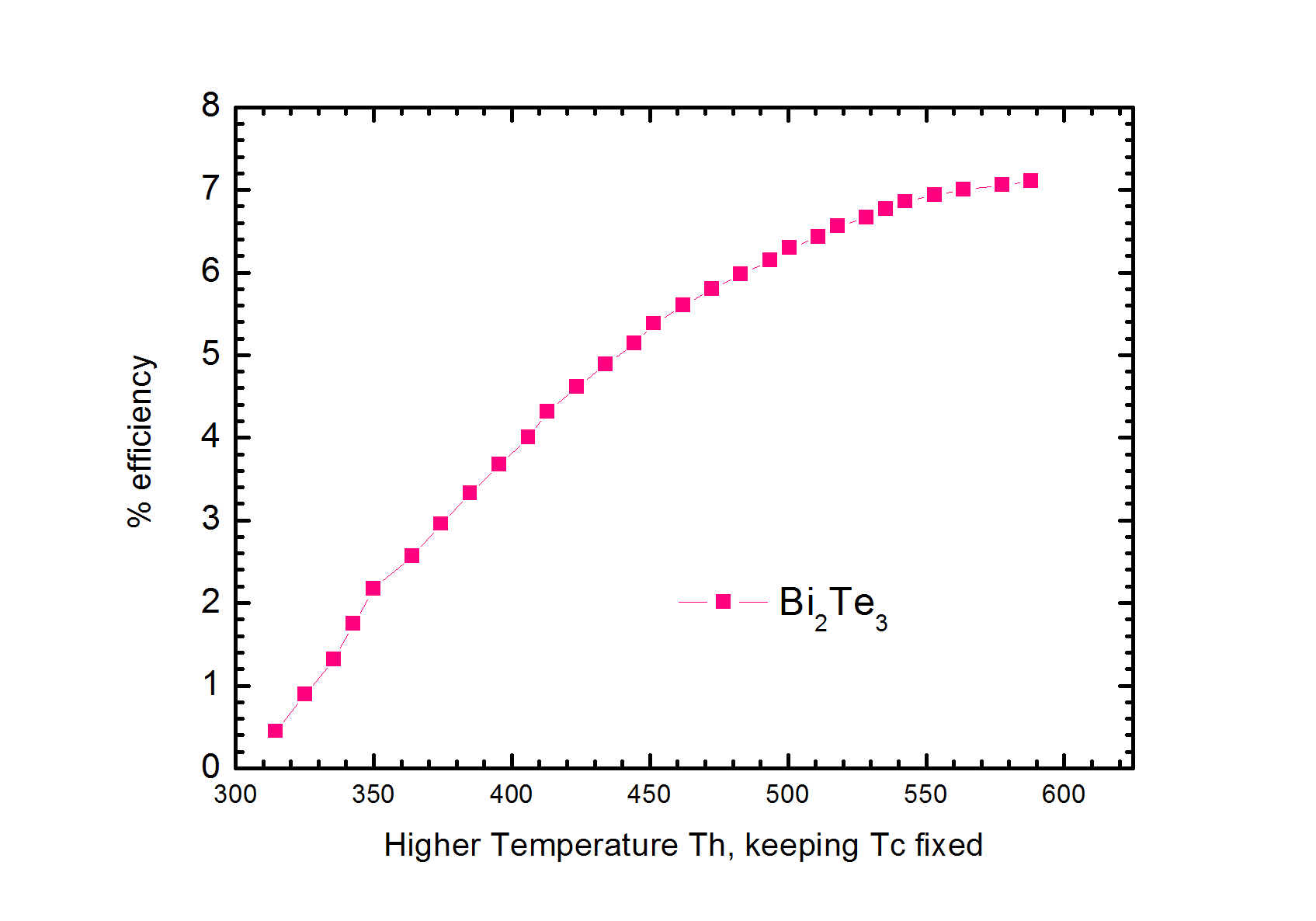}
\caption{showing efficiency v/s temperature difference through sample, keeping lower temperature fixed at 310 K}
\end{figure}
Another point to be noted that the efficiency is increased with decreasing rate with temperature and became almost constant at $T_{h} = 580 K$. This will help to judge the upper limit of temperature for efficient designing of TEG.\par The fairly good agreement between our calculated results with the experimental results obtained from TEG suggests the utility of our methods. Like, in predicting the efficiency of TEG under various condition by using material properties like S, $\sigma$, $\lambda$ on which figure of merit $z\bar{T}$ depends through Eq.(3). Since the temperature dependent data of S, $\sigma$, $\lambda$ for various materials are available in the literature, so this method provides a cheap and efficient way for evaluating the efficiency of a TEG made up of a particular material. We have curves for TEG efficiency made up of different materials, which will help in selecting a proper thermoelectric material for a specific application.\par The applicability of TEG made up of either $Bi_2Te_3$ or $TAGS$ or $SiGe$ material for different temperature range\cite{snyder2008complex} will be justified by installing in an automobile. In the automobile sector enormous amount of energy is being wasted out through exhaust port\cite{meisner}. Useful energy extracted through waste heat can be used as to supply power to the auxiliary need of automobile, which will reduce fuel consumption. In an automobile, upper temperature is governed by the type of engine installed in that vehicle (other parameters such as velocity, road condition, engine power, etc., constant). So, exhaust port temperature will vary from low, medium, and high for LMV (light motor vehicle), LTV (light transport vehicle), and HMV (heavy motor vehicle), respectively. In our case, the temperature of the hot end will be decided by the exhaust temperature of a particular type of vehicle. Thus, here we have estimated the temperature dependent efficiency of TEG with hot end kept at fixed temperature by varying cold end temperature, which is presented in Fig.3. So, we can easily find out the efficiency of TEG between fixed $T_{h}$ and variable $T_{c}$. It is important to note that in X-axis, we have taken the temperature difference between hot and cold end. For calculating the absolute temperature of the cold end, we have to subtract the temperature difference from the temperature of hot end.
\begin{figure}[ht]
\centering
\includegraphics[width=3.5 in]{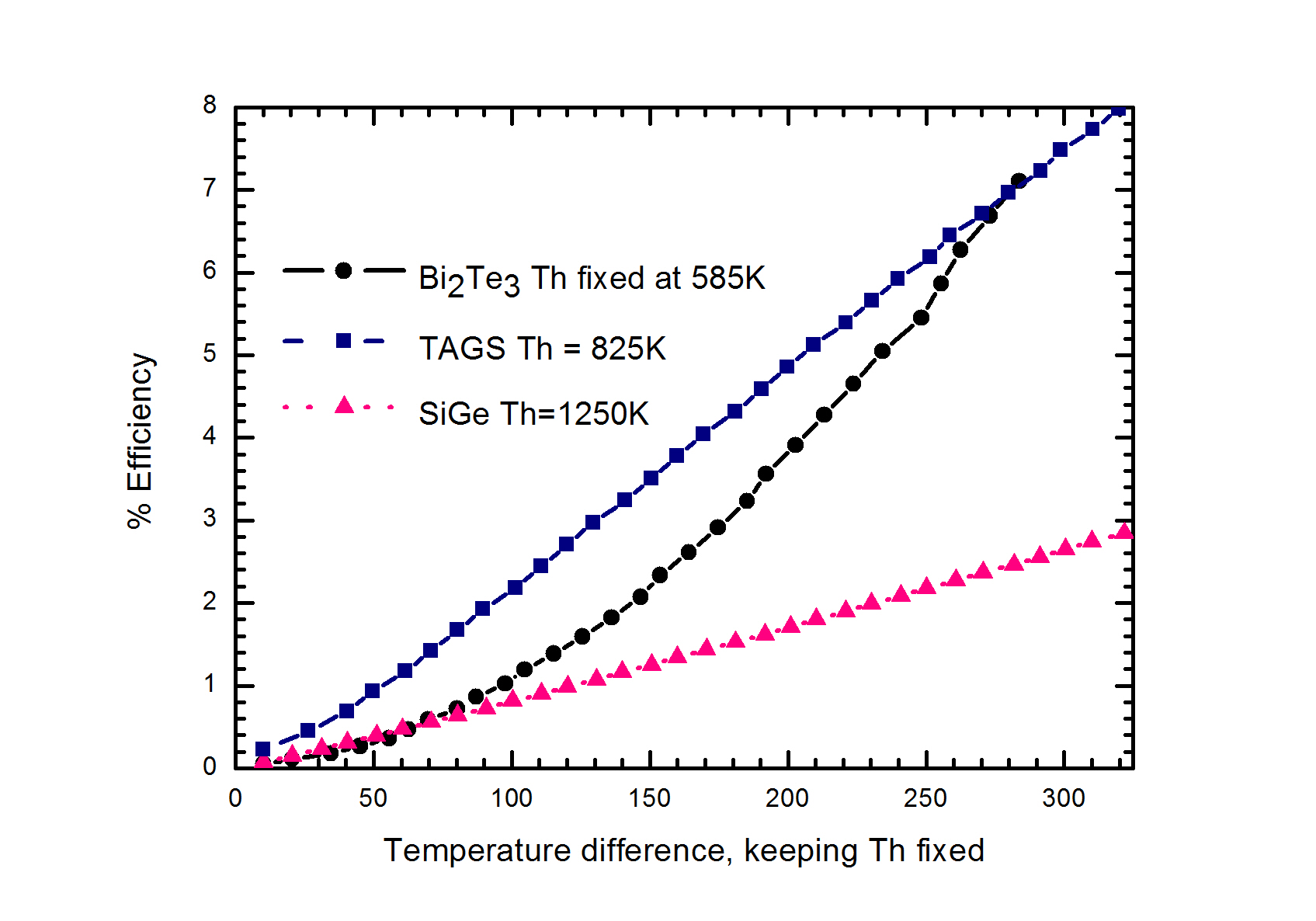}
\caption{variation of efficiency with varying temperature difference in different materials}
\end{figure}
We know that hot end temperature is fixed and cold side temperature is surrounding dependent, so for different conditions, there will be different cold side temperature. Suppose, we took a typical example of a cylindrical sample of $Bi_2Te_3$ having 0.5 cm diameter and 2 cm long, convective heat transfer coefficient as 5 $W/(m^{2}K)$. Considering, conductive as well as convective loss through the sample, we found out cold end temperature as 492 K, keeping hot end temperature fixed at 585 K. Similarly, only by doubling the length of the sample, we found out $T_c = 388 K$. Using Fig.3, we can tell about efficiency, which is 1\% and 4\%, respectively for two cases. Similarly, we can take $TAGS$ as a thermoelectric material for TEG within a typical temperature range of 700 K - 830 K. This gives 3\% efficiency. Suppose for a temperature range of 900 K - 1260 K, we are taking $SiGe$ as a thermoelectric material for TEG, which gives 3.4\%. Consequently, there will be different temperature differences between two ends of TEG for different lengths and correspondingly we can find out efficiency for that temperature difference. Since these curves can be used to calculate the efficiency of TEG directly for a particular temperature difference, so it may prove to be very useful for design engineers. Simultaneously to judge the best possible temperature range for maximizing efficiency of a TEG.
\section{CONCLUSION}
Calculation of TEG efficiency involves a great challenge because of temperature dependent material's properties. Hence, we have formulated a method to calculate the efficiency of TEG for power generation. Theoretically calculated results are very close to the experimentally reported results. By using the fundamental law of thermodynamics for a segmented system, the calculation of TEG efficiency is done. For simplifying our calculation, we applied the one-dimensional model for calculating efficiency. We took $Bi_2Te_3$, $TAGS$, and $SiGe$ as thermoelectric materials for TEG and calculated efficiencies come out to be 7.1\%, 8.94\%, and 3.5\%, respectively. This approach enabled us to directly calculate the efficiency of TEG for a particular temperature difference. We have also calculated the progressive change in efficiency by keeping one end temperature fixed with another end varying. Automobile's design engineer can use our results to fabricate efficient TEG. The basic understanding of efficiency calculation will give us some clue for material selection in designing of TEG, consequently, our time and money get saved.

\bibliographystyle{unsrt}

\end{document}